# Analysis of geospatial behaviour of visitors of urban gardens: is positioning via smartphones a valid solution?

Francesco Pirotti[1,2][0000-0002-4796-6406], Marco Piragnolo[1,2][0000-0003-0948-1691], Alberto Guarnieri[1,2][0000-0002-9483-289X], Marco Boscaro[1], Raffaele Cavalli[1][0000-0002-2704-7744]

[1] TESAF Department, University of Padova, Viale dell'Università, 16 - 35020 Legnaro (PD)
[2] CIRGEO Interdepartmental Research Center in Geomatics, University of Padova, Viale dell'Università, 16 - 35020 Legnaro (PD)

**Abstract.** Tracking locations is practical and speditive with smartphones, as they are omnipresent devices, relatively cheap, and have the necessary sensors for positioning and networking integrated in the same box. Nowadays recent models have GNSS antennas capable of receiving multiple constellations. In the proposed work we test the hypothesis that GNSS positions directly recorded by smartphones can be a valid solution for spatial analysis of people's behaviour in an urban garden. Particular behaviours can be linked to therapeutic spots that promote health and well-being of visitors. Three parts are reported: (i) assessment of the accuracy of the positions relative to a reference track, (ii) implementation of a framework for automating transmission and processing of the location information, (iii) analysis of preferred spots via spatial analytics. Different devices were used to survey at different times and with different methods, i.e. in the pocket of the owner or on a rigid frame. Accuracy was estimated using distance of each located point to the reference track, and precision was estimated with static multiple measures. A chat-bot through the Telegram application was implemented to allow users to send their data to a centralized computing environment thus automating the spatial analysis. Results report a horizontal accuracy below ~2.3 m at 95% confidence level, without significant difference between surveys, and very little differences between devices. GNSS-only and assisted navigation with telephone cells also did not show significant difference. Autocorrelation of the residuals over time and space showed strong consistency of the residuals, thus proving a valid solution for spatial analysis of walking behaviour.

## 1 Introduction

Higher accuracy geolocation solutions are increasingly making their way in mass-market sectors as devices increase their ability to perform low-end mapping and surveying activities [1]. Global navigation satellite systems (GNSS) are now served by multiple constellations. Multiple constellations mean that there are more satellites thus a higher probability of having an unobstructed line of sight (LOS) between the receiver and the space vehicle (SV). They also provide a higher probability of an improved satellite configuration, thus lower values of position dilution of precision (PDOP). GNSS receivers in smartphones are now capable of collecting information from multiple constellations. This capability is becoming more common also in lower cost smartphones,that can now receive United States' Global Positioning System - GPS (GPS), Global'naya Navigatsionnaya Sputnikovaya Sistema - GLONASS (GLO), Europe's Galileo (GAL), Japan's Quasi-Zenith - QZSS (QZS), China's BeiDou (CMP), and India's Indian Regional Navigational Satellite System - IRNSS (IRN).

Satellite navigation system data are shared through several standard formats. The GNSS chipset in the smartphone can provide coordinates in the NMEA (National Marine Electronics Association) protocol, which comes in different versions. NMEA 0183 is the most common format to this day, providing communication of position and information on the quality of satellite geometry. NMEA files provide position calculated through the Standard Positioning Service (SPS) or, if the chipset allows, real time differential corrections. In general, an SPS receiver can provide position information with an error of less than 10 meters. In this particular case the NMEA data was used.

Investigations using low-cost and smartphone-based receivers have already been extensively carried out. There is a high interest in understanding the possibility of using such devices for surveying and monitoring [2], both in real time kinematic and also with post-processing of observation data, even online as was tested in [3]. Pirazzi et al. [4] tested GPS+GAL single frequency antenna in a smartphone in different scenarios, finding metric accuracy in dynamic surveying and decimetric accuracy in static surveying. Dabove and Di Pietra [5] tested two smartphones and an external receiver, finding decimetric accuracies with centimetric precision in real time if a reference station is available. Dabove in [6] also tested GNSS NMEA acquisitions from two smartphones in two scenarios, one scenario providing mostly fixed solutions, and one with only floating solutions and reasonably multipath error sources. Results of comparing with reference positions from a total station showed horizontal deviations respectively ~3 m ~5 m. Lately also dual-frequency GNSS receivers (L1/E1/ and L5/E5 signals) have become available in few models of smartphones, which have been analysed for performance in [7] showing an improvement from ~5 m to ~1 m. Positions recorded in time also allow to record velocities, and Android smartphone (Xiaomi Mi 8) was tested also in this sense, finding root mean square errors (RMSE) of a few millimeters per second in east and north directions [8].

## 2  Materials and methods

Different devices are tested under controlled conditions in different scenarios. In particular locations under different types of canopy cover were tested, as the accuracy reduction due to canopy cover is the main object of investigation.

### 2.1  Study area

The tests were carried out in the premises of the University of Padova, Villa Revedin Bolasco (VRB), located in the city of Castelfranco Veneto (province of Treviso).

**Historical background.** The area, where today the villa and the garden stand, was historically owned by noble families, Tempesta, Morosini and since 1509, to the Venetian patrician family Corner which built their residence. Italian garden decorated with statues, the work of Orazio Marinali (1643-1720) was called "The Paradise". The garden was designed as a Romantic English garden designed by Meduna and other famous landscape designers such as Francesco Bagnara and Marc Guignon from France. Today, the historical garden extends over an area of 7.63 hectares (Fig. 1), and it counts over 1,000 trees. Owning to the many elements that are involved, a specific database was also implemented for management of maps, profiles and other topographical information [9].

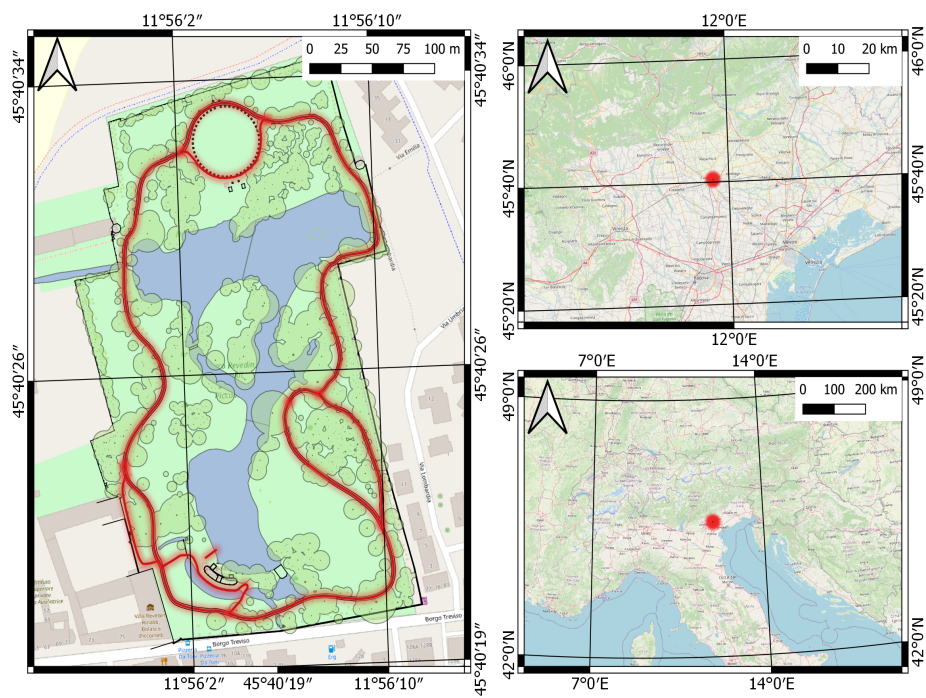

**Fig. 1.** Study area, Villa Revedin Bolasco. The red line is the track used as reference. Geographic coordinates refer to the WGS84 datum.

## 2.2 Surveys

In this note the term "survey" is used to indicate the collection of coordinates over a defined time-lapse. In the reported tests the "Rinex ON" application was used [10, 11]. Rinex ON was also used in other investigations [12–15], and allows to collect data from the NMEA protocol. Rinex ON also stores GNSS observations and navigation messages in Receiver Independent Exchange (RINEX) format that can be used for further post-processing, but this will be done in a future investigation on this topic and is not in the scope of this paper.

Five smartphones were used for repeated surveys. Surveys were carried out, individually and, in one specific day, 3 february 2021, with all smartphones tested simultaneously. On that day three surveys were done with all smartphones in the same frame (see Fig. 2). The frame keeps a consistent relative position of all four smartphones during movement and allows more rigorous comparisons. The frame was positioned at a height of 1.8 m to make sure that it is above the heads of surveyors, thus avoiding signal obstruction by body parts (e.g. head).

**Table 1.** Date and type of test survey. Constellation references (**Sat**): GPS (1), GLONASS (2), Galileo (3), QZSS (4), BeiDou (5), IRNSS(6)

| Smartphone | Sat | Smartphone | Sat |
|---|---|---|---|
| Blackview - BV4900 | 1,2,3,5 | Xiaomi - Redmi Note 8T | 1,2,5 |
| Motorola - Moto E (4) Plus | 1 | Xiaomi - Redmi Note 9S | 1,2,3,5,6 |
| Samsung - SM-A515F | 1,2,3,5 | | |

A total of 19 surveys were carried out. On the 3rd of february 2021, more smartphones were tested simultaneously in three surveys. The first two surveys on that specific date were done by fixing the smartphones on a frame positioned above the head (see Fig. 2), and the third and last survey of that day was done keeping the smartphones in the pocket of each owner.

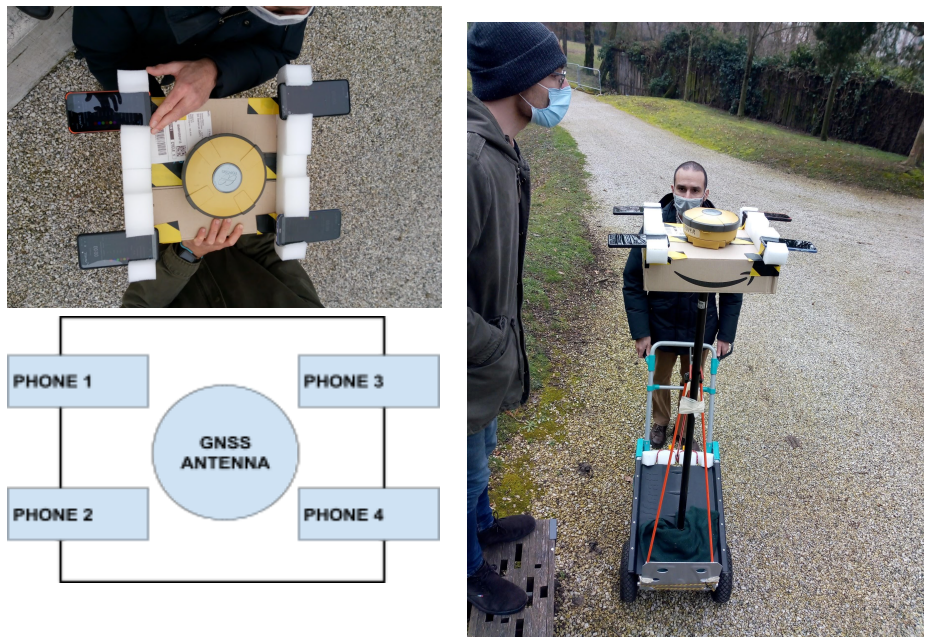

**Fig. 2.** Experimental setup (left) and survey using carriage.

Each survey lasted about 26 minutes: 20 minutes of mobile measures, and three minutes of static measurements at the beginning and three minutes at the end. Starting times of the campaign were respectively at 11:31 AM, 12:08 AM and 12:32 AM (see Table 2). Each round-trip along the track collected around 1500 coordinates.

The following sample design was applied for this day: (i) smartphones were positioned in a specific spot and Rinex ON application was started; (ii) smartphones were left to collect measures statically for three minutes; (iii) the smartphones were carried through navigation at constant walking speed on a closed-ring trip on the middle of the track in Fig. 1; (iv) the navigation stopped along the track in two predetermined spots for about 10 seconds; (v) the smartphones were positioned back to the initial spot and were left there for three minutes again to statically collect data; (v) Rinex ON was stopped and the NMEA file sent to the server via Telegram through the T-BOT (see next section).

**Table 2.** Survey at single date (2021-02-03) with multiple smartphones.

| ID | Survey | Smartphone model | ID | Survey | Smartphone model |
|---|---|---|---|---|---|
| 1 | 11:31:00 | Xiaomi - Redmi Note 8T | 7 | 12:08:00 | Xiaomi - Redmi Note 8T |
| 2 | 11:31:00 | Xiaomi - Redmi Note 9S | 8 | 12:32:00 | Xiaomi - Redmi Note 8T |
| 3 | 11:31:00 | Samsung - SM-A515F | 9 | 12:32:00 | Xiaomi - Redmi Note 9S |
| 4 | 12:08:00 | Blackview - BV4900 | 10 | 12:32:00 | Blackview - BV4900 |
| 5 | 12:08:00 | Samsung - SM-A515F | 11 | 12:32:00 | Motorola - Moto E (4) Plus |
| 6 | 12:08:00 | Xiaomi - Redmi Note 9S | | | |

### 2.3 NMEA file - transmission and analysis via T-BOT

To automate the process of collecting the data from the surveys and streamlining them on a workflow for processing, a specific system was deployed. Transmission over the internet was implemented by creating a Telegram chat-bot (T-BOT) and collecting the files to a database through server-side. Analysis was successively carried out in the R CRAN (R) environment. **Fig. 3** depicts the data flow and the points below summarize it.

1. the user downloads to the smartphone an app that supports recording GNSS data and provides results in the NMEA format (Rinex ON in our tests).
2. The user starts the app and collects positions, just like in our tests.
3. The user ends the survey and shares the results through the T-BOT.
4. R functions processes data to detect preferred spots and analyse behaviour of visitors.

In our case the last point included processing accuracy metrics because the users walked on the reference track. The NMEA files provided through Rinex ON also contain the date and time of the start of the survey and also the model of the smartphone in the header. This allows to automatically keep track of the individual survey, because the T-BOT receiving the NMEA file provides information about the user sending the file. Therefore it is possible to record information on the user, the smartphone model and the date/time of the survey.

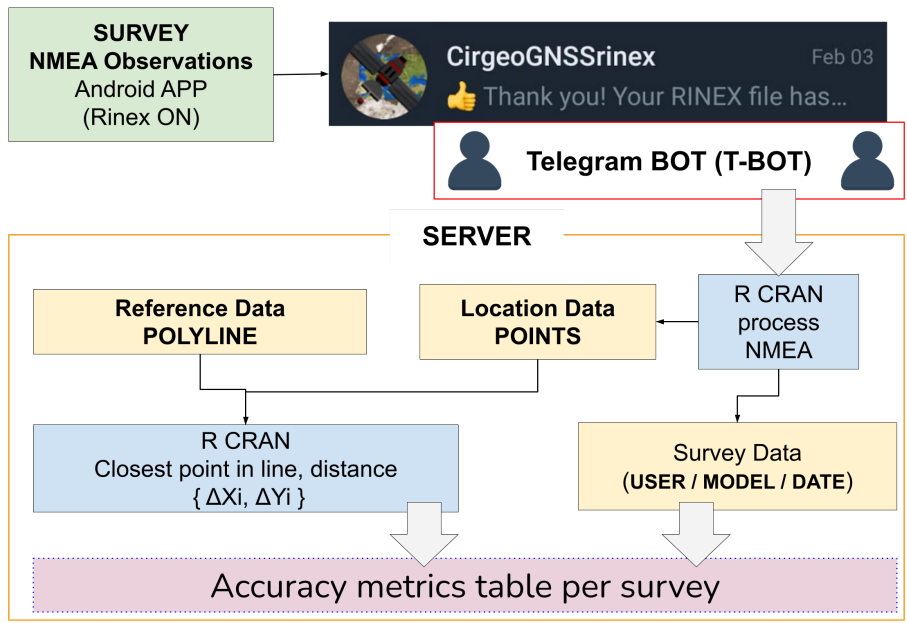

**Fig. 3.** Semi automatic workflow for ingesting and processing NMEA data from the survey via a ChatBOT developed in Telegram application.

### 2.4 Analysis of geolocation quality

The points were recorded every second always along a reference track (Fig. 1), so in each survey campaign they can be considered a time series, i.e. a sequence of measurements of the same variables made over time. Points are recorded as geographic coordinates in latitude and longitude in the WGS84 coordinate reference system (CRS) frame. For further processing geospatial coordinates were transformed to the projected UTM CRS in zone 32 ETRS89 (EPSG code 25832).

The reference track in Fig. 1 was surveyed using a total station instrument, with monumentation of points for each station for a closed polygonization. The polyline consisted of points surveyed on the track every ~1 m from the stations. The closure of the polyline allowed it to minimize the errors by a least-squares approach and assess the measurement error, that was < 0.19 m. The assessment was done by moving along this track. It is therefore reasonable to say that the distance between measured points and this track is a proxy of accuracy. Being the track circular, we can also assess, to a certain degree, any directional bias regarding the residuals.

**Accuracy.** Estimated accuracy was calculated for each survey by measuring the residuals for each recorded coordinate, $x_0$ and $y_0$. Residuals were calculated by considering as real coordinate the point on the polyline of the reference track (see Fig. 1) that was closest to $x_0$ and $y_0$ with the following formula:

$$x = \frac{b(bx_0 - ay_0) - ac}{\sqrt{a^2 + b^2}} \quad y = \frac{b(-bx_0 + ay_0) - bc}{\sqrt{a^2 + b^2}} \quad (1)$$

where x and y are the coordinate values of the point on the line which is closest to $x_0$ and $y_0$, and a, b and c are constants where a and b cannot both be zero.

It must be noted that this is an empirical way to assess accuracy, as the closest point on the reference line is not necessarily the real coordinate. Another way to look at this shortcoming is that the directional component of the residuals that corresponds to the orientation of the segment will not be detected. This is an empirical method that is sensitive to the components of the residuals that are perpendicular to the direction of the segment considered of the polyline, i.e. the nearest one to the considered point, and is not sensible to errors in the same direction. This is a necessary drawback to assess a mobile survey, but it is acceptable because the goal is not to assess accuracy of the single point, but provide an idea of the average accuracy for a full survey. Being the track a circular path, sensibility to errors in all directions will also be assessed. This will be further discussed in the next sections.

**Precision**. The replicability of the measure was estimated by leaving the smartphones positioned statically for three minutes at the beginning and at the end of each survey, as described in section 2.2. The standard deviation from the mean was calculated in the east and north directions to provide an estimation of precision.

**Temporal and spatial error correlation.** Something is temporally correlated when its values are similar over different moments in time. Temporal and spatial autocorrelation of GNSS residuals have been investigated and found to be significant [16–19]. Previous work has shown that accuracy changes at a defined rate, e.g. in the work of Olynik [21] 5 cm in 50 seconds was observed. This is an advantage in the application that is proposed in this work, because the target is to detect when visitors move and when they stop or slow down; in other words the determination of velocity of the movement of the visitor. Velocity is calculated by distance between position in two successive intervals t0 and t1 (one second at the sampling rate of the surveys). If the error between t0 and t1 is very similar, then the difference of the two positions will partly remove the error. As noted in Ranacher et al. [20] "if there is a strong autocorrelation between any two consecutive position estimates, they have very similar errors. These errors cancel out when average speed, distance or direction is calculated along the trajectory.".

## 3     Results and discussion

The objective of this work is to verify the feasibility of using personal devices (smartphones) for determining the behaviour of visitors in an urban garden context. In particular to understand if the coordinates recorded at a rate of 1 Hz through smartphones are precise enough to understand which are the favored spots where visitors linger longer periods. The point to consider is the following: are GNSS NMEA data suitable for implementing a system where smartphones can transmit positions to a central analysis system that will be able to process data to find hot-spots and thus analyse behaviour? In this context factors that can decrease the quality of the geolocation data are vegetation canopy cover, sensitivity loss of GNSS antenna due to the position (smartphone will realistically be carried by the user in the hand or in the pocket or purse) and, to a lesser extent, multi-path from buildings. First accuracy and precision are reported in the following section, then discussion on the feasibility of spatial analysis focused on user behaviour is reported.

### 3.1     Quality of geolocation

**Accuracy.** The accuracy is reported in two plots Fig. 4 and Fig. 5 below.

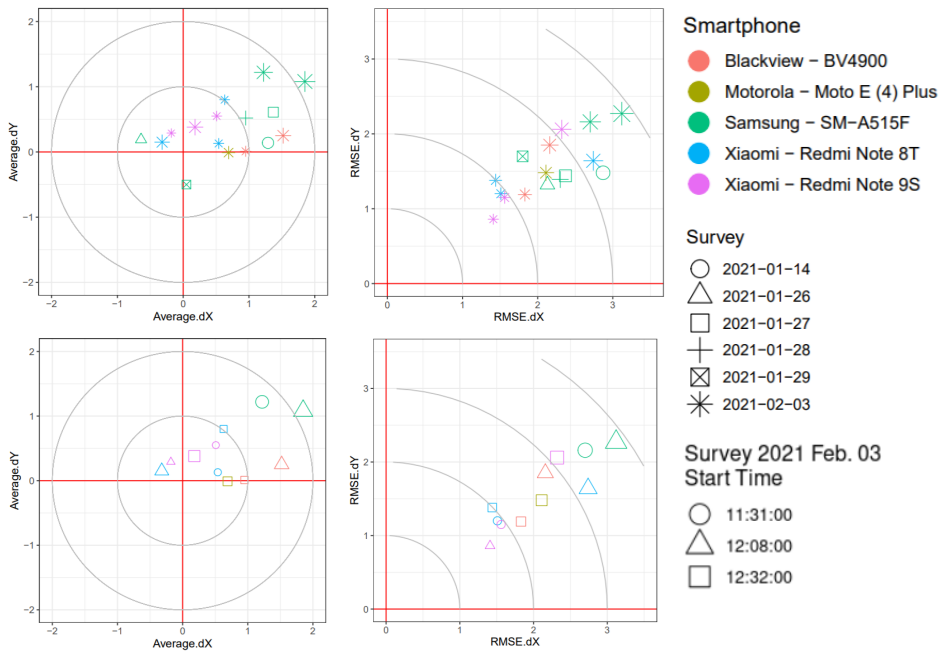

**Fig. 4.** Distribution of average of residuals (left column), and RMSE of residuals (middle column) and legends (right column).

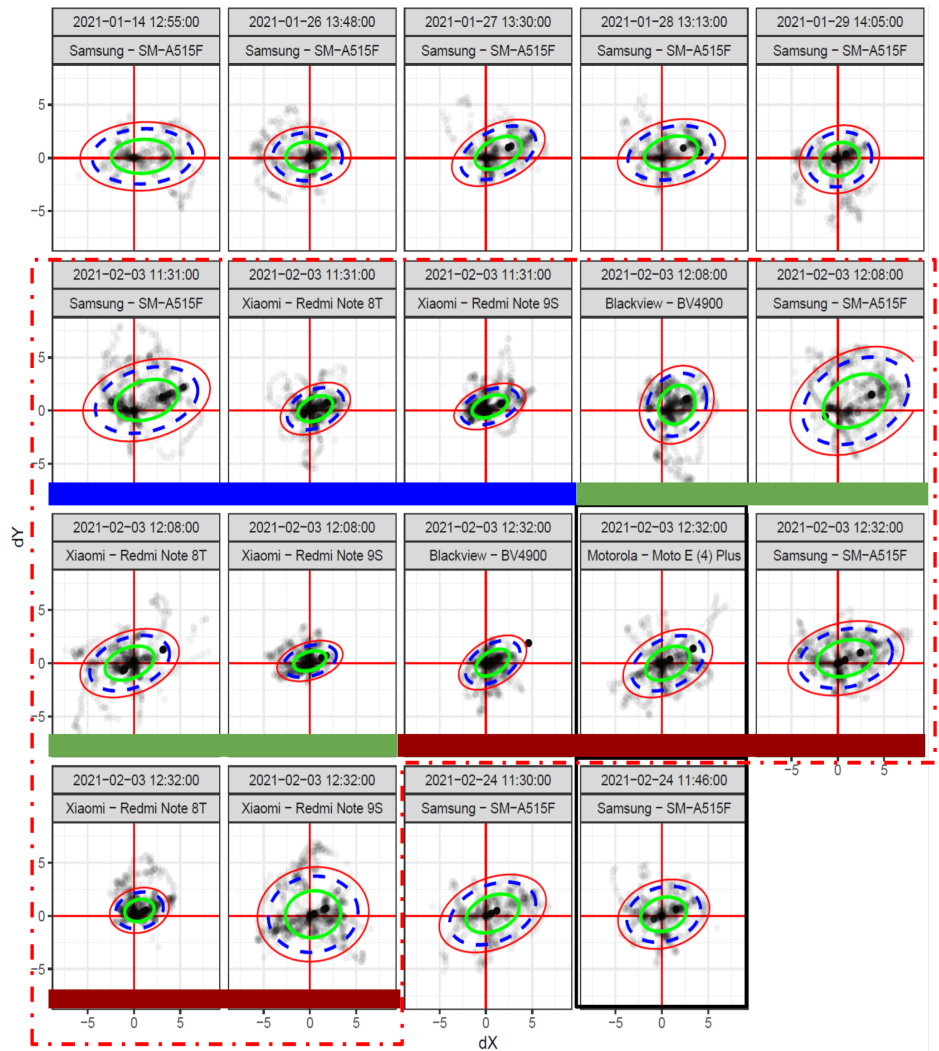

**Fig. 5.** Distribution of residuals. Green blue and red ellipses respectively indicate 68% 95% and 99% confidence intervals and denote estimated accuracies. Boxes with black borders indicate surveys without phone cell signal (aero mode). Red dashed line indicates the concurrent survey campaign on 3$^{rd}$ of february 2021 with multiple smartphones. Thick bottom ribbon color indicates the date of simultaneous surveys.

Again it is important to note that the estimation of accuracy here was done using a reference track, thus a linear geometry which underestimates the residual component with a cosine law of the direction of the closest segment to each considered point. Nevertheless, the number of measurements and the circular shape of the track allows to sample all directions. Results do show a slight anisotropy in the east-west direction, as many ellipses shown in Fig. 4 show directionality. This is likely due to the fact that the track is oriented in the north-south direction, thus most readings will be compared with segments that have a north-south direction, thus diminishing that component of the error and weighting more the east-west component.

**Precision.** The three minutes of static measures with the smartphones at the beginning and the end of the survey allowed us to calculate the replicability of the measure, also known as the precision. It must be noted first the intrinsic precision of the NMEA format. The NMEA format provides coordinates as *ddmm.mmmmm* where "d" is degrees and "m" is minute - thus 4 factional digits of the minute. The weight of the least significant digit is therefore ~0.186 m and ~0.131 m respectively for latitude and longitude values for our position. The standard deviation ($\sigma$) of 100 values of coordinates while the smartphone is statically positioned is shown in Table 3, with colored cells highlighting the lowest precisions. It can be seen that all values are below 1 m.

**Table 3.** Precisions (1$\sigma$) from static positioning during 2021-02-03

| Survey | Sensor | START | | END | |
|---|---|---|---|---|---|
| | | X (m) | Y(m) | X(m) | Y(m) |
| 2021-02-03 11:31 | Samsung - SM-A515F | 0.824 | 0.499 | 0.000 | 0.000 |
| 2021-02-03 11:31 | Xiaomi - Redmi Note 8T | 0.133 | 0.025 | 0.028 | 0.104 |
| 2021-02-03 11:31 | Xiaomi - Redmi Note 9S | 0.154 | 0.935 | 0.014 | 0.043 |
| 2021-02-03 12:08 | Blackview - BV4900 | 0.001 | 0.000 | 0.074 | 0.260 |
| 2021-02-03 12:08 | Samsung - SM-A515F | 0.145 | 0.422 | 0.127 | 0.861 |
| 2021-02-03 12:08 | Xiaomi - Redmi Note 8T | 0.094 | 0.125 | 0.166 | 0.601 |
| 2021-02-03 12:08 | Xiaomi - Redmi Note 9S | 0.076 | 0.058 | 0.275 | 0.943 |
| 2021-02-03 12:32 | Blackview - BV4900 | 0.000 | 0.000 | 0.000 | 0.000 |
| 2021-02-03 12:32 | Motorola - Moto E (4) Plus | 0.783 | 0.630 | 0.000 | 0.000 |
| 2021-02-03 12:32 | Samsung - SM-A515F | 0.335 | 0.393 | 0.026 | 0.001 |
| 2021-02-03 12:32 | Xiaomi - Redmi Note 8T | 0.275 | 0.305 | 0.033 | 0.015 |
| 2021-02-03 12:32 | Xiaomi - Redmi Note 9S | 0.016 | 0.025 | 0.056 | 0.113 |

A more detailed description is provided in Fig. 6 which shows that the distribution of the measures is quite irregular, with many cases having the exact same coordinate with respect to the least significant digit. For example in the 12:32 survey the BV4900 has $\sigma$=0.00 meaning that all one hundred values had the exact same values, thus perfect precision (always limited to the intrinsic precision of the NMEA format). It is also important to note that temporal autocorrelation, which is discussed in the next section, will provide very similar residuals during the three minutes, and therefore if the smartphones are positioned for a longer time, the precision will decrease as the drift in the error is caught by the larger temporal window.

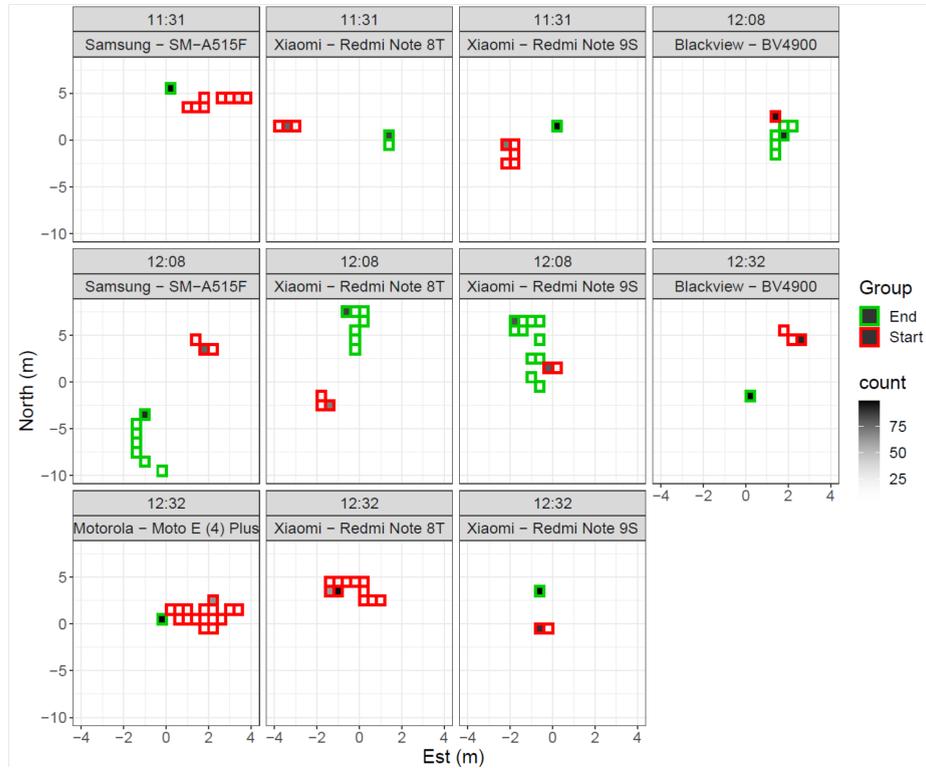

**Fig. 6.** Distribution of coordinates during the static measures at the start and the end of the surveys. Dark dots mean that all measures are inside that bin, light color inside the bins means very few measures.

### 3.2 Autocorrelation

As mentioned it is proven in previous investigations that the residuals are not random over time, but change gradually, thus change at a certain rate. It is important for the goal of this investigation, because the application of using positions is to detect movement and speed, and not absolute location information. The coefficient of correlation between two values in a time series is defined by the autocorrelation function (ACF). Fig. 7 shows the results of this metric over most surveys done in March 03 2021. The plots report a lag time of 2 minutes and show that on average all surveys are still significantly correlated on a 30 seconds lag, consistently over the X and Y components. It means that the residual can be modelled using the previous residuals in time. These results are coherent with the ones from [16–19] where GNSS residuals have been investigated and found to be significant, as well as from the work of Olynik [21] where a change of 5 cm in 50 seconds was observed.

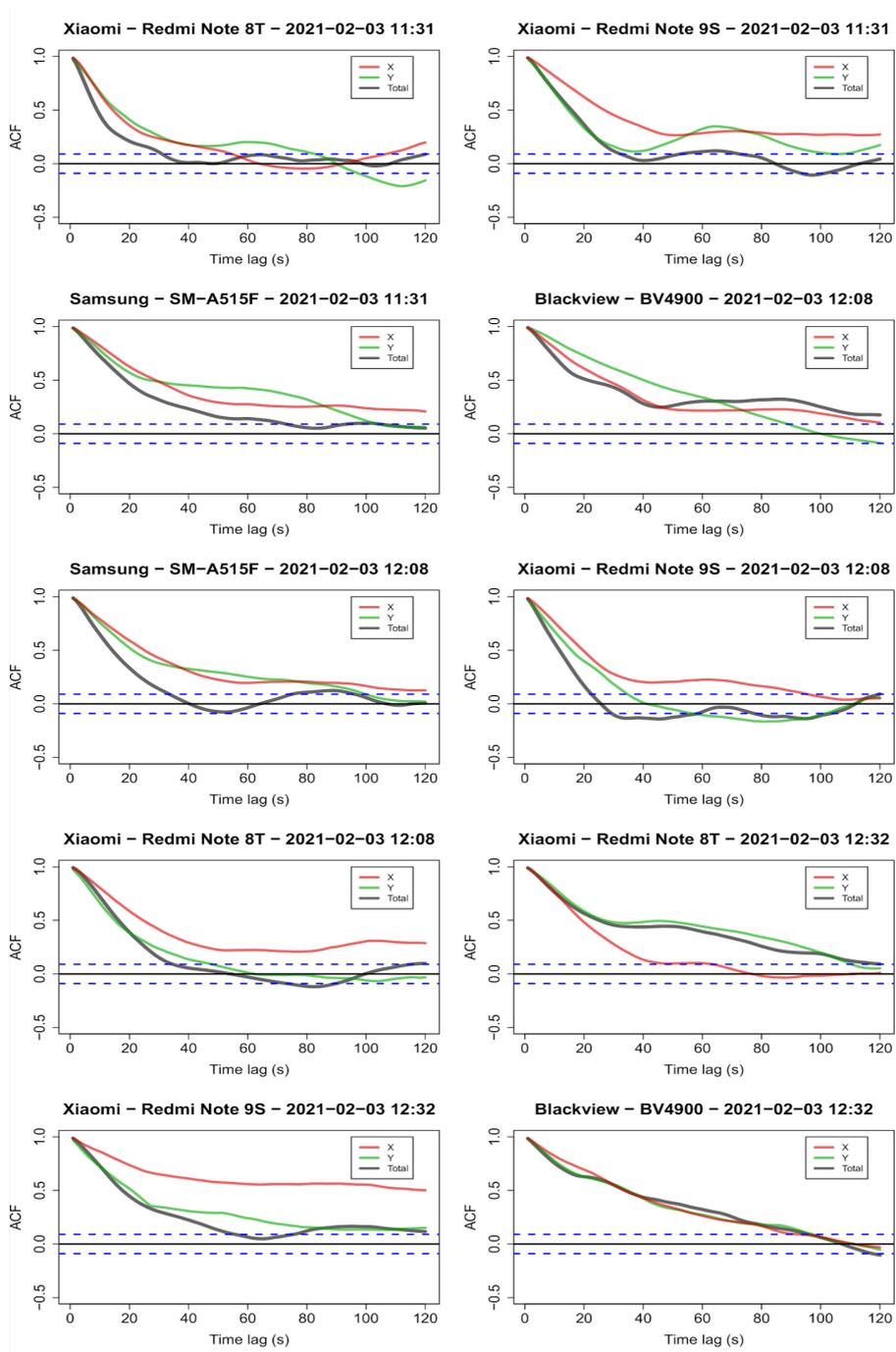

**Fig. 7.** Autocorrelations values on a lag time up to 2 minutes.

### 3.3 Clustering "hotspots"

The final result is the detection of the three spots where the simultaneous surveys stopped for around 10 seconds. The spots correspond to benches along the track. These are depicted in Fig. 8. Four spots are detected because the bottom left one corresponds to the start/end position, where the smartphones measured statically for three minutes and thus were identified as a stopping spot. As mentioned in the previous section, autocorrelation allows to abate the effect of the error when calculating the speed of the movement by simply the first derivative of distance over time. For best results it was observed that a low-pass median filter over a 5 second time window helps to remove noise, thus false positive stops. For false positives it is meant that two or three successive points might be very close in space due to the effect of the error, even if in reality there was movement that caused the points to be apart. Considering only two-three points would provide a false signal of the smartphone stopping. To avoid this a low-pass filter removes these false signals.

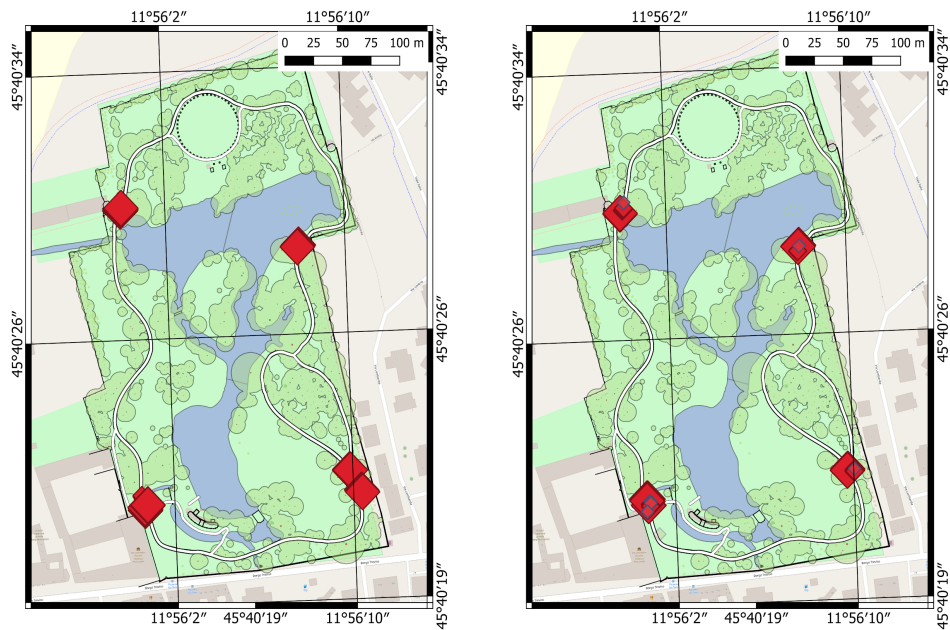

**Fig. 8.** Detection of "preferred" spots from filtering geolocations which did not change significantly for 10 seconds or more. Left is the first survey (rigid-frame) and right is the last survey (hand-held) in date 03 March 2021.

## 4 Conclusions

It is well known that the rigorous estimation of accuracy requires to compare a measured value with another value that was measured with an instrument with several orders of magnitude higher accuracy. In this work a rougher estimation of accuracy is enough for the intended focus, which is to test how well a smartphone allows us to understand the behaviour of a visitor in terms of relative position in time. In other words we want to know if it is possible, carrying a smartphone in a garden with varying cover conditions, to understand if a person slows down in certain points or speeds up on other areas which maybe are not as pleasant. Results show that positions have overall errors that range a couple of meters from the reference, but also that this error remains constant enough to be able to define speed accurately enough to detect changes in movement. It is also reported that data can be streamed via a chat-bot to automate processing and remove time-consuming file transfers via other means. Another part worth noting is that all smartphones tested provided similar results, without significant differences, even if they had different capabilities in terms of how many constellations they received and model of GNSS antenna.

### 4.1 Further development

Here results were reported regarding the suitability of smartphones for traking positions in a specific environment, i.e. a different types of canopy cover. Villa Revedin Bolasco is also the site of investigations regarding the well-being of visitors, in particular elderly people, as part of the VARCITIES EU H2020-EU.3.5.2 project. It is expected that voluntary visitors use their own smartphone, if it meets the criteria for GNSS receivers, or be provided with a smartphone during the visit. Participation will be on a voluntary basis. The T-BOT will automatically collect and store data in the server, which will also trigger a post-processing and analysis workflow.

Future work will look more closely on the use of post-processing methods on the RINEX files which were also collected during the survey with Rinex ON. It is planned to collect more data with other applications to investigate improved solutions with respect to only using data from the NMEA protocol of the GNSS chipset. Another further future work will be pushing further automation providing a service that automatically post-processes RINEX observations and answers back through the Telegram BOT chat with corrected positions.

## Acknowledgements

This work was supported by the VARCITIES project, Grant Agreement number: 869505 — VARCITIES — H2020-SC5-2018-2019-2020 / H2020-SC5-2019-2. Processing was carried out by support from the Hyperearths project - ISREDI7542

15